\documentclass[twocolumn,printnumbers,amsmath,amssymb,showpacs]{revtex4}
\usepackage{graphicx}% Include figure files
\usepackage{dcolumn}% Align table columns on decimal point
\usepackage{bm}% bold math
\begin{document}
%\preprint{APS/123-QED}

\title{Energy transport in jammed sphere packings}

\date{\today} 

\author{Ning Xu$^{1,2}$} 
\author{Vincenzo Vitelli$^1$} 
\author{Matthieu Wyart$^3$}
\author{Andrea J. Liu$^1$}
\author{Sidney R. Nagel$^2$}

\affiliation{$^1$Department of Physics and Astronomy, University of
  Pennsylvania, Philadelphia PA, 19104; $^2$The James Frank Institute, The
  University of Chicago, Chicago IL, 60637; $^3$HSEAS, Harvard University,
  Cambridge MA, 02138}

\begin{abstract}
  We calculate the normal modes of vibration in jammed sphere packings to
  obtain the energy diffusivity, a spectral measure of transport.  At the boson peak frequency, we find an Ioffe-Regel crossover from a diffusivity that drops rapidly with frequency to one that is nearly frequency-independent.  This crossover frequency shifts to zero as the system is decompressed towards the jamming transition, providing unambiguous evidence of a regime in frequency of nearly constant diffusivity.   Such a regime, postulated to exist in glasses to explain the temperature dependence of the thermal conductivity, therefore appears to arise from properties of the jamming transition.
 
\end{abstract}

\maketitle

Zero-temperature soft-sphere models inspired by foams and granular media have
given insight not only into the geometry of hard-sphere
packings~\cite{O'Hern03,Donev05,Silbert06} but also into the physics of the
low-energy excitations in glasses~\cite{Leo1,Xu07}.  In particular, a model system of
frictionless spheres interacting with finite-ranged repulsions exhibits a
jamming transition (Point J) at a density corresponding to random
close-packing of hard spheres~\cite{O'Hern03}.  At the transition, the
coordination number jumps~\cite{Durian95,O'Hern03} from zero to the minimum
value required for mechanical stability, the ``isostatic"
value~\cite{alexander}, and there is a plateau in the density of vibrational
states that extends down to zero frequency~\cite{Leo1}.  Upon compression,
this plateau persists but only above a characteristic frequency, $\omega^*$,
that increases with density.  The modes in the plateau region have been shown
to arise from zero-frequency vibrational modes at the isostatic
transition~\cite{matthieu}.  These anomalous modes are in excess of the Debye
prediction and are directly connected~\cite{matthieu2,Xu07} to excess
vibrational modes in glasses, known as the ``boson
peak"~\cite{Bosonpeak,taraskin,gurevich}.  However, it is not 
clear how these modes contribute to heat conduction.

In this paper, we investigate thermal transport as a function of compression
in jammed sphere packings.  At the jamming threshold, we find that all delocalized modes
transport heat with a low diffusivity nearly independent of frequency, in contrast to
ordinary solids in which sound modes transport heat ballistically with a
diverging diffusivity in the long-wavelength limit.  The behavior at Point J
is reminiscent of many amorphous solids, which unlike crystals display a thermal conductivity 
that rises monotonically with temperature $T$~\cite{Pohl02}.
This property has
been posited to arise from a frequency regime of small, constant
diffusivity~\cite{Kittel,Sheng91,Allen93}.  We show that this regime
can originate from the vibrational spectrum at Point J.  Upon compression,
the low-diffusivity modes persist, but only above a crossover frequency corresponding to the  frequency of the boson peak, $\omega^*$~\cite{Parshin,Tanaka};
below $\omega^*$, the spectrum is dominated by transverse plane waves.

Our model~\cite{O'Hern03,Leo1} is a 50/50 mixture of frictionless spherical
particles with a diameter ratio of 1.4.  Particles $i$ and $j$ interact in
three dimensions via a one-sided harmonic potential: $
  U(r_{ij})=\frac{\epsilon}{2}(1-r_{ij}/\sigma_{ij})^2$ when the distance
between their centers, $r_{ij}$, is less than the sum of their radii,
$\sigma_{ij}$ and zero otherwise.  Jammed packings at temperature
$T=0$ are obtained by conjugate-gradient energy minimization.  We study
systems of $250 \le N \le 10,000$ particles with periodic boundary
  conditions in all directions.  The packing fraction at the onset of
jamming, $\phi_c$, is characterized by the onset of a nonzero pressure.  We
determine $\phi_{c}$ and obtain $T=0$ configurations at controlled $\Delta
\phi \equiv \phi-\phi_c$ as in Ref. \cite{Leo1}.  For each configuration, we
diagonalize its Hessian matrix, whose $m^{th}$ eigenvalue is the squared
frequency, $\omega_m^{2}$, of the orthonormal eigenmode described by the
displacement $\vec e_m (j)$ of each particle $j$. We used the package
ARPACK to handle large systems~\cite{ARPACK} . The particle mass, $M$,
interaction energy, $\epsilon$, and diameter of the smaller particle,
$\sigma$, are set to unity.  The frequency is in units of
$\sqrt{\epsilon/M\sigma^2}$.

We also study an ``unstressed" model in which we use energy-minimized
configurations obtained from the previous model, and replace the interaction
potential ${ U(r_{ij})}$ between each pair of overlapping particles with an
unstretched spring with the same stiffness, $U^{\prime \prime}(r_{ij})$.
Because all springs are unstretched, there are no forces between particles in
their equilibrium positions so that stable configurations for the stressed
system are also stable in the unstressed one.  This corresponds to dropping
terms depending on $U^\prime$ in the Hessian~\cite{alexander,matthieu}.

For a strongly scattering system, a diffusive description of energy transport
can be more useful than one in terms of ballistic propagation with a very short
mean free path.  Therefore instead of calculating the
thermal conductivity, $\kappa(T)$, directly using molecular
dynamics~\cite{Jund,sheng}, we calculate the thermal diffusivity,
$d(\omega_m)$ for vibrational mode $m$ ~\cite{Sheng91,Allen93}.  $\kappa(T)$ can be expressed in terms
of $d(\omega_m)$ and the heat capacity
$C(\omega_m)$~\cite{John,Sheng91,Allen93}:
\begin{equation}
 \kappa(T)=\frac{1}{V}\sum_m C(\omega_m) d(\omega_m)  
=\frac{1}{V}\int_{0}^{\infty} {\rm d}\omega D(\omega)  C(\omega) d(\omega) ,
\label{eq:conductivity}
\end{equation}
where the sum runs over all vibrational modes $m$, $V$ is the volume of
  the system, and $D(\omega)\equiv \sum_{m} \delta(\omega_m - \omega)$ is the
density of vibrational states.  Thus, in Eq.~\ref{eq:conductivity},
$C(\omega)= k_B {(\beta \hbar\omega)}^2 e^{\beta\hbar\omega}/{(e^{\beta
    \hbar\omega}-1)^2}$ (where $\beta \equiv 1/k_B T$ and $k_B$ is the Boltzmann constant) depends on $T$.  It characterizes the heat carried at
frequency $\omega$, while $d(\omega)$, which has units of $\sigma
\sqrt{\epsilon/M}$, is a $T$-independent scattering function.

The physical meaning of the diffusivity is best illustrated operationally.
Consider a wavepacket narrowly peaked at $\omega$ and localized at
$\vec r$ at time $t=0$.  Over time, the wavepacket spreads out and
can be characterized by a time-independent diffusivity given by the square of
the width of the wavepacket at time $t$, divided by $t$ \cite{Sheng91}.  In a
weakly-scattering system, $d(\omega)=c \ell(\omega)/3$, where $c$ is the sound speed
and $\ell(\omega)$ the phonon mean-free path.

We follow Allen and Feldman~\cite{Allen93} to calculate 
$d(\omega)$ in terms of the normal modes of a given configuration, using the
Kubo-Greenwood formula for the thermal conductivity as the response to a
temperature gradient that couples different modes.  We use \cite{Allen93}
\begin{equation}
d(\omega)=\frac{\pi}{12 M^2 \omega ^2} \int_0^\infty {\rm d}\omega^\prime D(\omega^\prime) \frac{(\omega +\omega^\prime)^2}{4\omega \omega^\prime}|\vec{\Sigma}(\omega,\omega^\prime)|^2 \delta(\omega-\omega^\prime) 
\label{eq:DCdiffusivity}
\end{equation}
where the vector heat-flux matrix elements are
\begin{equation}
|\vec \Sigma(\omega,\omega^\prime)|^2=\frac{\sum_{mn} |\vec \Sigma_{mn}|^2 \delta (\omega-\omega_m)\delta(\omega^\prime-\omega_n)}{D(\omega) D(\omega^\prime)}
\label{eq:Sigmadef}
\end{equation}
where $m$ and $n$ index the vibrational modes.

For a finite system the modes are discrete. We calculate the matrix elements
$\vec{\Sigma}_{mn}$ from the Hessian $H_{\alpha\beta}^{ij}$ and its
$m^{th}$ normalized eigenvector $e_{m}(i; \alpha)$ \cite{Allen93} via
\begin{equation}
\vec{\Sigma}_{mn}= \sum_{i j,\alpha \beta} (\vec{r}_{i}-\vec{r}_{j}) e_{m}(i; \alpha) \!\!\!\! \quad H_{\alpha\beta}^{ij} \!\!\!\! \quad e_{n}(j; \beta),
\label{eq:Sigmamn}
\end{equation}
where $\{i,j\}$ and $\{\alpha,\beta\}$ label particles and their Cartesian
coordinates respectively.

In a finite system, the delta function in Eq.~\ref{eq:DCdiffusivity} must be
replaced by a representation with nonzero width, $\eta$.  We use~\cite{Allen93} $
g(\omega_m-\omega_n,\eta)=\eta/[\pi((\omega_{m}-\omega_{n})^2 + \eta^2)]$ and
define
\begin{equation}
d(\omega_m,\eta,N)\equiv\frac{\pi}{12 M^2 \omega_m ^2} \sum_{n\neq m} \frac{(\omega_m +\omega_n)^2}{4\omega_m \omega_n} |\vec{\Sigma}_{mn}|^2 g(\omega_n-\omega_m,\eta).
\label{eq:limitdef}
\end{equation}
We set $\eta=\gamma\delta\omega$, where $\gamma>1$ and
  $\delta\omega$ is the average spacing between successive modes. The desired $d(\omega)$ is then $d(\omega,\eta,N)$ in the double limit $N
\rightarrow \infty$, $\eta \rightarrow 0^+$.
%%%%%%%%%%%%%%%%%%%%%%%%%%%%%%%%%%%%%%%%%%%%%%%%%%%
\begin{figure}
\includegraphics[width=0.38\textwidth]{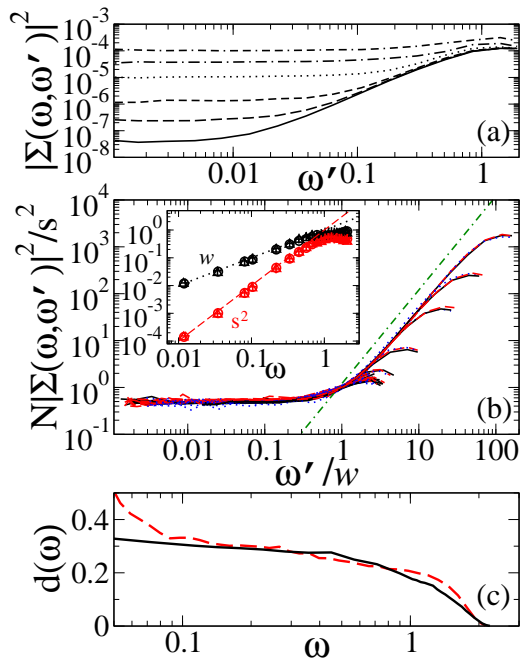}
\caption{\label{fig1} (Color Online) Diffusivity just above the jamming 
  transition at $\Delta \phi=10^{-6}$.  (a) Heat-flux matrix elements
  $|\Sigma(\omega, \omega' )|^2$ plotted versus $\omega'$ at $N=2000$ for
  $\omega=0.012$ (solid), $0.035$ (long dashed), $0.08$ (short dashed), $0.10$
  (dotted), $0.22$ (dot-dashed), and $0.68$ (dot-dash-dashed). (b) Scaling
  plot showing collapse of $|\Sigma(\omega,\omega')|^2$ at $N=2000$ (black
  solid), $1000$ (red dashed), and $500$ (blue dotted) with scale factors
  $s^2$ and $w$.  The green dot-dashed line
indicates a power-law slope of $2$.  Inset: Scale factors $s^2$ (red symbols) and
  $w$ (black symbols) versus $\omega$.  We find $s^2 \propto \omega^2$ (red
  dashed line) and $w \propto \omega$ (black dotted line) except at high
  $\omega$.  (c) Plot of $d(\omega,\eta=0.002,N=2000)$ defined in Eq.
  (\ref{eq:limitdef}) (dashed).  Solid line: predicted $d(\omega)$ for $N \rightarrow \infty$.}
  \end{figure}
%%%%%%%%%%%%%%%%%%%%%%%%%%%%%%%%%%%%%%%%%%%%%%%%%%%

We now use Eqs.~[2-5] to calculate the diffusivity.  Our goal is to extract
$d(\omega)$ for an infinite system so we must confront finite-size effects.
We will show that $|\vec \Sigma|^2$ has a particularly simple form at the
jamming threshold, enabling us to determine the $N \rightarrow \infty$
behavior.

Figure \ref{fig1}(a) shows the heat-flux matrix elements $|\vec
\Sigma(\omega,\omega^\prime)|^2$ defined in Eq.~\ref{eq:Sigmadef} for packings
at $\phi-\phi_c=10^{-6}$ for different values of $\omega$ versus
$\omega^\prime$.  Fig.~\ref{fig1}(b) shows that all the curves can be
collapsed for different system sizes, $N$, and
frequencies, $\omega$, except at high $\omega^\prime$ where the modes become
localized~\cite{Grest,Leo2,zorana}.  The inset to Fig.~\ref{fig1}(b) shows
that the scale factors for the collapse satisfy $s^2=\omega^2$ and $w=\omega$,
respectively, except at high frequencies where localization sets in.  Note
that the scaling collapse demonstrates that the only noticeable system-size
dependence is a prefactor of $1/N$.  Since for large $N$, the density of
states scales as $N$~\cite{Leo1}, Eq.~\ref{eq:DCdiffusivity} therefore yields
a well-defined diffusivity in the $N \rightarrow \infty$ limit (solid curve in Fig.~\ref{fig1}(c)).

Note that the collapse in Fig.~\ref{fig1}(b) implies that $|\vec
\Sigma(\omega,\omega)|^2\propto\omega^2/N$ at low frequencies.  This scaling
arises when overlap with nearby modes, described by Eq. (\ref{eq:Sigmamn}), is
small and independent of frequency and when modes are spatially
uncorrelated~\cite{Vincenzo}.  Thus Eq.~\ref{eq:DCdiffusivity} implies that
$d(\omega)\propto D(\omega)$ at low $\omega$.  At Point J, because the density of states is
nearly constant down to $\omega=0$~\cite{Leo1}, the diffusivity is nearly
constant as well.  (The small slope of $d(\omega)$ with frequency is due to
the slight frequency dependence of $D(\omega)$.)  These results show that
although the low-frequency modes are extended at the jamming
threshold~\cite{Leo2}, they do not behave like plane waves and the usual
divergence of diffusivity is completely suppressed.

Over most of the frequency range, this $N \rightarrow \infty$ prediction
agrees very well with the dashed curve in Fig.~\ref{fig1}(c), which shows
$d(\omega,\eta=0.002,N=2000)$ for a finite system.  However, at low
frequency $d(\omega,\eta,N)$ exhibits an upturn; this upturn is a finite-size
artifact that can be shown from Eqs.~[2-5] to scale as $\omega^{-3}$ with a
prefactor that vanishes as $N \rightarrow \infty$, $\eta \rightarrow
0$~\cite{Vincenzo}.

%%%%%%%%%%%%%%%%%%%%%%%%%%%%%%%%%%%%%%%%%%%%%%%%%%
\begin{figure}
\includegraphics[width=0.38\textwidth]{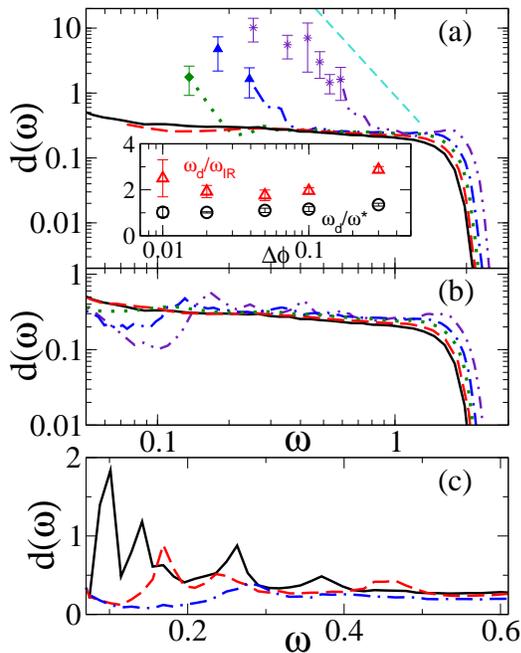}
\caption{\label{diffcomp} (Color Online) Diffusivity versus compression.
  $d(\omega, \eta=0.002,N=2000)$ for the (a) unstressed and (b) stressed
  systems at $\Delta \phi=10^{-6}$ (solid black), $0.01$ (red dashed), $0.05$
  (green dotted), $0.1$ (blue dot-dashed), and $0.3$ (purple dot-dot-dashed).
  In (a) the cyan dashed line indicates a power-law of $\omega^{-4}$ and the
  closed symbols indicate the degenerate sets of discrete plane-wave modes in
  our finite system.  Inset: The ratios $\omega_d/\omega^*$ (open circles) and
  $\omega_d/\omega_{IR}$ (open triangles) versus $\Delta \phi$.  (c)
  $d(\omega,\eta=0.004,N)$ for the stressed system at $\Delta \phi=0.5$ for
  $N=10,000$ (black solid), $2000$ (red dashed), and $500$ (blue dot-dashed).}
\end{figure}

%%%%%%%%%%%%%%%%%%%%%%%%%%%%%%%%%%%%%%%%%%%%%%%%%%%
%%%%%%%%%%%%%%%%%%%%%%%%%%%%%%%%%%%%%%%%%%%%%%%%%%

We now study diffusivity as the system is compressed beyond the jamming
threshold.  We begin with the unstressed model where the data are particularly
simple to interpret.  In this case, the system is always held at zero
pressure, so that increasing $\Delta \phi \equiv \phi-\phi_c$ corresponds to
increasing only the average coordination number of the network of interacting
particles.  At all compressions, Fig.~\ref{diffcomp}(a) shows that
$d(\omega,\eta=0.002,N=2000)$ vanishes at high $\omega$ for localized modes.
At low frequencies, there is the 
upturn due to finite-size effects discussed above; this upturn lies at frequencies below those shown in
Fig.~\ref{diffcomp}.  At intermediate frequencies, the
diffusivity decreases rapidly with increasing $\omega$ until it reaches a small constant value $d_0$ for $\omega>\omega_d$.  Below
$\omega_d$, the modes are discrete due to the finite system size, as indicated by the discrete points in Fig.~\ref{diffcomp}(a).
The mode frequencies correspond to $\omega=c_T k_n$, where $c_T$ is the
transverse sound speed and $k_n$ are the lowest allowed wavevectors. The calculated diffusivity below $\omega_d$ decreases sharply with increasing $\omega$
as expected for scattering of plane waves~\cite{John}.  Thus, $\omega_d$
marks the crossover from transverse plane waves to a small, nearly
constant diffusivity.

The frequency $\omega_d$ can be understood as the Ioffe-Regel
crossover from weak to strong scattering of transverse
modes~\cite{Ioffe}.  For $\omega<\omega_d$, the transverse plane-waves obey
$\omega=c_Tk$, where $k$ is the wavevector.  As $\omega$ approaches
$\omega_d$, we have shown $d(\omega) = c_T\ell/3 \rightarrow d_0$, where
$\ell$ is the phonon scattering mean free path.  The Ioffe-Regel criterion, $k
\ell \approx 1$, predicts a crossover frequency near $\omega_{IR} \equiv
c_T^2/3d_0$.  For our harmonic system, the transverse speed $c_T \propto
\Delta \phi^{0.25}$ \cite{O'Hern03} so that $\omega_{IR} \propto \Delta \phi^{0.5}$.  The inset to Fig.~\ref{diffcomp} indeed shows
for transverse modes, $\omega_d/\omega_{IR} \approx 2$ over our
range of compression, consistent with the Ioffe-Regel criterion.  

Fig.~\ref{diffcomp}(a) (inset) also shows the ratio of $\omega_d$ to the boson peak frequency, $\omega^*$, defined as the onset frequency for the plateau in the
density of states, which was previously shown to scale as $\Delta \phi^{0.5}$ \cite{Leo1,Xu07}.  The ratio $\omega_d/\omega^*$ is constant over a wide range of
$\Delta\phi$.  Studies of silica~\cite{Vacher,Horbach} and several disordered models~\cite{Parshin,Tanaka} observe that the boson peak frequency and Ioffe-Regel crossover frequency agree within a factor of order unity.
In our system, this relation is unambiguous because both frequencies shift together as $\Delta \phi$ is varied.
Our results also indicate that at Point J, where $\omega^*=0$, the diffusivity should remain nearly flat down to zero frequency, as shown independently in Fig.~\ref{fig1}(c).  Therefore, the modes above $\omega_d$ ({\it i.e.,} those with
constant diffusivity) can be identified as the anomalous modes that derive
from soft modes at the isostatic point~\cite{matthieu,matthieu2}.  

While the unstressed system may be appropriate for systems where the
coordination at threshold exceeds the isostatic value (such as
frictional systems~\cite{Leiden}), we are also interested in systems where
inter-particle forces increase with $\Delta \phi$.  These forces lower both the sound speed $c_T$ which controls the Ioffe-Regel crossover frequency~\cite{O'Hern03} and the frequencies of modes in the plateau~\cite{matthieu}.  Finite-size effects, which cut off plane waves at low $\omega$, are therefore more
obstructive in stressed systems.  

Fig.~\ref{diffcomp}(b) shows that there is no discernible change in the
diffusivity of the stressed system over the range $10^{-6} \le \Delta \phi \le 10^{-2}$.  Above
$\Delta \phi \approx 10^{-2}$, structure develops at intermediate frequencies.
Each peak can be identified with one of the first few allowed wavevectors for
longitudinal or transverse modes~\cite{Vincenzo}.  
Fig.~\ref{diffcomp}(c) shows that as $N$ increases, the plane-wave peaks shift to lower
frequencies and grow closer together, as expected, so that for high enough
$N$, peaks in a given frequency range will merge into a smooth curve as peak
widths exceed their spacing.  We therefore conclude that this structure will
disappear in the infinite-size limit when the only observable plane-wave peaks
will be shifted to zero frequency.

Fig.~\ref{diffcomp}(c) also shows that the peak heights increase with $N$ and
decreasing $\omega$ at low frequency.  This suggests that the diffusivity
rises smoothly with decreasing $\omega$ at frequencies below some $\omega_d$
in the thermodynamic limit, and has a constant value $d_0$ above $\omega_d$.
In the unstressed case, we found plane-wave behavior below $\omega_d \sim
\omega_{IR} = c_T^2/3d_0$.  In the stressed case, we speculate that $\omega_d$
might be similarly defined and should also increase with $\Delta \phi^{0.5}$
since $c_T \propto \Delta \phi^{0.25}$ for the stressed as well as the
unstressed case~\cite{O'Hern03}.  
%%Our results shed light on experiments on sound propagation in compressed granular packings, which reveal that a coherent {\it ballistic} pulse coexists with a specklelike {\it diffusive} signal \cite{Liu92, Jia99}.  This is consistent with our observation of both plane-wavelike modes and poorly conducting ones at low frequency.

At the jamming threshold, we can calculate
the thermal conductivity, $\kappa(T)$, from Eq.~\ref{eq:conductivity}, and obtain a finite answer even within the harmonic approximation because the diffusivity does not diverge at low $\omega$.   By setting $d(\omega)$ and $D(\omega)$ to be constant up to the localization threshold, we find $\kappa \propto T$ up to the Debye temperature.  At high temperatures, $\kappa$ saturates when all the modes are excited. 

One of the most striking differences between heat conduction in ordered and disordered structures is that 
the thermal conductivity $\kappa$ of crystalline materials first rises with increasing $T$ but eventually drops due to phonon-phonon scattering~\cite{Pohl02}, while $\kappa$ for glasses increases monotonically in $T$.  This perplexing property of glasses has been explained heuristically 
by assuming that phonons are scattered so strongly by disorder that transport becomes diffusive, with a frequency regime of small, constant
thermal diffusivity.  In that case, the thermal conductivity simply
increases with the heat capacity according to Eq. \ref{eq:conductivity} \cite{Kittel,Sheng91,Allen93}.  In our finite-sized unstressed
systems, we see clear evidence for such a regime of nearly constant diffusivity, and find that its frequency onset, given by the Ioffe-Regel crossover for transverse phonons, increases with compression.  For the compressed system with stress,
our results are much less clear in the low-$\omega$ regime because of
finite-size effects.  However, above some intermediate frequency, we again
find a constant diffusivity and constant density of states, which lead to the
rise in the thermal conductivity as in the unstressed case.
  
In earlier work, we showed that the vibrational spectra of model systems such
as the Lennard-Jones glass could be understood in terms of jammed
sphere packings \cite{Xu07}.  We now find that these packings capture some of
the crucial physics invoked to explain the temperature dependence of the thermal
conductivity--a crossover from low-$\omega$ transverse phonons to excess vibrational modes of
nearly constant diffusivity~\cite{Tanaka}.
The physical origin of the diffusive modes lies in the behavior of packings
at the jamming threshold, where the Ioffe-Regel crossover frequency vanishes.  Upon compression,
the flat diffusivity shifts to higher frequencies but does not
disappear.  Thus, compressed sphere packings are a useful starting point
for understanding energy transport in glasses.
  
%\begin{acknowledgments}
We thank W. Ellenbroek, R. D. Kamien, T. C. Lubensky, Y. Shokef and T. A.
Witten for helpful discussions.  This work was supported by DE-FG02-05ER46199
(AL, NX and VV), DE-FG02-03ER46088 (SN and NX) , NSF-DMR05-47230 (VV), and
NSF-DMR-0213745 (SN).

%\end{acknowledgments}

\end{document}